\documentclass{article}
\usepackage{spconf,amsmath,graphicx}
\usepackage{amsfonts}

\usepackage[font=small,labelfont=bf]{caption}
\title{Multi-channel Narrow-Band Deep Speech Separation with Full-band Permutation Invariant Training}
\name{Changsheng Quan ${ }^{1,2}$, Xiaofei Li ${ }^{2,*}$\thanks{* corresponding author}}
\address{${}^1$ Zhejiang University, Hangzhou, China\\
${}^2$ Westlake University \& Westlake Institute for Advanced Study, Hangzhou, China}

\begin{document}
%
\maketitle
\begin{abstract}
This paper addresses the problem of multi-channel multi-speech separation based on deep learning techniques.
In the short time Fourier transform domain, we propose an end-to-end narrow-band network that directly takes as input the multi-channel mixture signals of one frequency, and outputs the separated signals of this frequency.
In narrow-band, the spatial information (or inter-channel difference) can well discriminate between speakers at different positions.
This information is intensively used in many narrow-band speech separation methods, such as beamforming and clustering of spatial vectors.
The proposed network is trained to learn a rule to automatically exploit this information and perform speech separation.
Such a rule should be valid for any frequency, thence the network is shared by all frequencies.
In addition, a full-band permutation invariant training criterion is proposed to solve the frequency permutation problem encountered by most narrow-band methods.
Experiments show that, by focusing on deeply learning the narrow-band information, the proposed method outperforms the oracle beamforming method and the state-of-the-art deep learning based method. 
\end{abstract}
\section{Introduction}
\label{sec:intro}
This work addresses the multi-channel speech source separation problem leveraging deep learning techniques. 
Traditional speech source separation methods are often designed in the short-time Fourier transform (STFT) domain.
In \cite{yilmaz_blind_2004}, the authors introduced the W-disjoint orthogonality assumption based on the time-frequency (TF) sparsity of speech, namely each TF bin of the mixed speech signals is assumed to be dominated by one speech source. Consequently, speech sources can be separated by clustering the TF bins. 
The authors of \cite{winter_map-based_2006} proposed to estimate the mixing matrix using hierarchical clustering of the TF-wise mixing vectors.
Beamforming conducts speech separation by applying spatial filtering.
The widely used minimum variance distortionless response (MVDR) beamformer preserves the speech with desired steering vector while suppresses others \cite{gannot_consolidated_2017}.
The recently proposed Guided Source Separation (GSS) \cite{boeddecker_front-end_2018} method combines the techniques of TF-bin clustering and MVDR beamforming, which achieves excellent speech separation performance, and thus is extensively adopted by the participants of the CHiME-6 multispeaker speech recognition challenge \cite{watanabe_chime-6_2020}.
The traditional methods mentioned above \cite{winter_map-based_2006, gannot_consolidated_2017, boeddecker_front-end_2018} are all performed in narrow-band, namely processing the STFT frequencies separately.
This leads to the well-known frequency permutation problem that the separated signals at different frequencies should be assigned to the same source.

Deep learning has been first used for single-channel speech separation by learning the spectral pattern of speech.
Deep clustering (DC) \cite{hershey_deep_2016} first learns an embedding for each TF bin, then uses k-means to cluster these embeddings and separate the TF bins.
Permutation invariant training (PIT) \cite{yu_permutation_2017} directly predicts the TF mask of each source, and then the separated signals can be obtained by multiplying the masks with the mixture signal.
PIT uses the minimum loss of all possible source permutations for training.
When multiple microphones are available, in addition to the spectral pattern of speech, the spatial information of speakers can be employed.
The single-channel DC method is extended to the multi-channel case in \cite{wang_multi-channel_2018} by integrating the spatial features, such as inter-channel phase difference, into the network input.
In \cite{gu_enhancing_2020}, it is proposed to automatically learn spatial features with neural network.
An end-to-end filter-and-sum network (FaSNet) \cite{luo_fasnet_2019} is proposed to directly predict the spatial filters from the time-domain mixture signals.
It is further combined with the transform-average-concatenate (TAC) paradigm in \cite{luo_end--end_2020}, which eventually achieves excellent speech separation performance.
Another popular technique is to estimate the beamformer parameters using the speech separation results of deep learning based methods, such as in \cite{ochiai_beam-tasnet_2020, heymann_blstm_2015}.

In this work, we propose an end-to-end narrow-band speech separation network.
A long short-term memory (LSTM) network is designed to take as input the STFT coefficients of multi-channel mixture signals for one frequency, and predict the STFT coefficients of multiple speech sources for the same frequency.
By analyzing the traditional narrow-band methods \cite{winter_map-based_2006, gannot_consolidated_2017, boeddecker_front-end_2018}, we can find that one STFT frequency includes rich informations to separate speech sources, such as the spectral sparsity of speech and the inter-channel differences of multiple sources.
The proposed network is trained to learn a function/rule to automatically exploit these informations, and to perform end-to-end narrow-band speech separation.
As is the case for traditional methods, one unique function/rule should be learned for all frequencies, thus the proposed network is designed to be shared by all frequencies. 

The proposed narrow-band method also suffers from the frequency permutation problem.
To solve this problem, the spectral correlation of adjacent frequencies, or the spatial consistency of multiple frequencies for the same speaker, can be used \cite{sawada_robust_2004, mazur_approach_2009}.
In this paper, inspired by utterance-level PIT (uPIT) \cite{kolbaek_multitalker_2017}, we propose a training criterion called full-band PIT (fPIT) to solve the frequency permutation problem.
It requires the network to output the separated signals of all frequencies belonging to the same speaker at the same output position.
This scheme can be implemented in an end-to-end manner within the training process.

This paper is a continuous work of our previous papers \cite{li_multichannel_2019, li_narrow-band_2019}, in which a narrow-band network was proposed for speech denoising by exploiting the differences between speech and noise, as speech is non-stationary and directional while noise is stationary and spatially diffuse.
Narrow-band speech separation is a different task, in the sense that it mainly exploits the inter-channel cues of different speakers.
Compared to the full-band multi-channel methods, such as FaSNet \cite{luo_fasnet_2019, luo_end--end_2020}, even though our proposed method does not learn the spectral pattern of speech at all, it employs a powerful network dedicated to fully leverage the narrow-band information.
Experiments show that the proposed method achieves better performance than FaSNet.
The code for our proposed method can be found at\footnote{https://github.com/Audio-WestlakeU/NBSS}.

\section{Method}
\label{sec:method}

\begin{figure}[tb]
  \centering
  \includegraphics[width=0.4\textwidth]{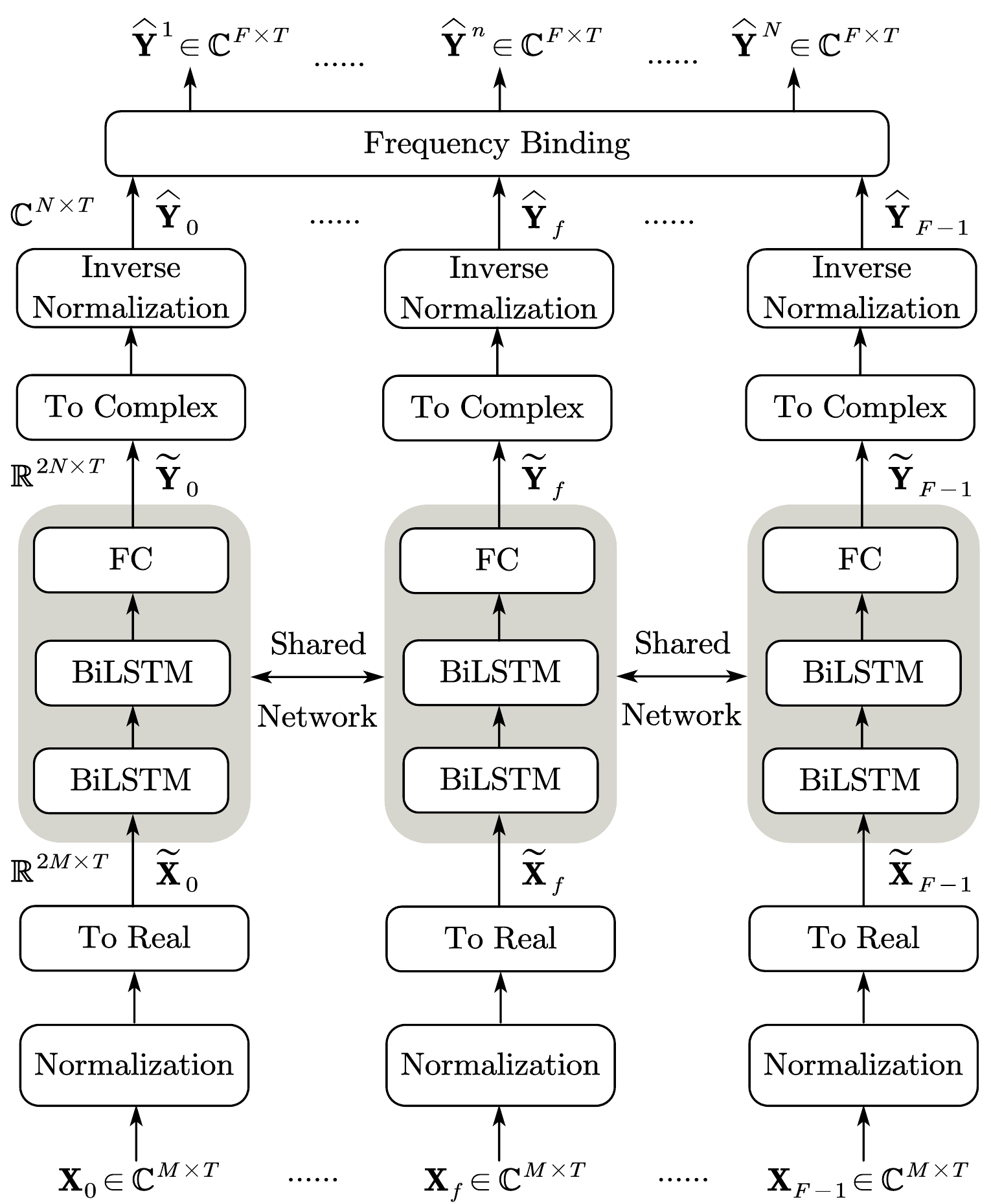}
  \vspace*{-3mm}
  \caption{The flowchart of narrow-band deep speech separation. Each frequency is processed individually. The dimension of each intermediate layer is on its leftmost, e.g.\ $\boldsymbol{\widetilde{\rm X}}_{f}\in  \mathbb{R}^{2M\times T}$.}
  \label{fig1}
\end{figure}

We consider multichannel signals in the STFT domain: ${\rm X}_{f,t}^{m} = \sum_{n=1}^N {\rm Y}_{f,t}^{n,m}$
, where $f$ $\in$ $\{0,...,F-1\}$, $t$ $\in$ $\{1, ..., T\}$, $m$ $\in$ $\{1,...,M\}$ and $n\in \{1,...N\}$ denote the indices of frequency, time frame, microphone channel and speaker, respectively,
${\rm X}_{f,t}^{m}$ and ${\rm Y}_{f,t}^{n,m}$ are the complex-valued STFT coefficients of the microphone signals and reverberant spatial image of speech sources, respectively.
This work focuses on the separation task, and our target is to recover the reverberant spatial image of multiple speakers at a reference channel, e.g.\ ${\rm Y}_{f,t}^{n,r}$, where $r$ denotes the index of reference channel.

\subsection{Narrow-band Deep Speech Separation}

The flowchart of the proposed method is shown in Fig.\ref{fig1}.
The basic idea is to perform speech separation independently for each frequency.

The STFT coefficient of multi-channel mixture signals are directly taken as the input feature of the network. For one TF bin, the STFT coefficients are concatenated along channels:
\begin{equation}
\mathbf{X}_{f,t} = [{\rm X}_{f,t}^1,\dots,{\rm X}_{f,t}^M]^{T} \in \mathbb{C}^{M\times 1}
\end{equation}
where $^{T}$ denotes vector transpose. Then, the time sequence of $\mathbf{X}_{f,t}$ for one frequency is taken as the input sequence of the RNN network:
\begin{equation}
\mathbf{X}_{f} = (\mathbf{X}_{f,1},\dots,\mathbf{X}_{f,T}) \in \mathbb{C}^{M\times T}
\end{equation}

To facilitate the network training, all input sequences are normalized to have the magnitude mean for the reference channel being one. The normalization is performed as $\mathbf{X}_{f}/ \overline{{\rm X}}_f$, where $\overline{{\rm X}}_{f} = \textstyle\sum_{t=1}^T |{\rm X}_{f,t}^r|/T$.
As the network can only process real numbers, the complex-valued input sequence should be converted into real-valued sequence.
This is done by simply replacing the complex number with its real part and imaginary part.
The real-valued sequence is denoted by $\widetilde{\mathbf{X}}_{f}\in R^{2M\times T}$.

The outputs of the network are the separated signals for this frequency.
In the literature, the magnitude spectra, TF mask or complex mask are often chosen as the training target \cite{wang_supervised_2018}.
However, it was demonstrated in our previous work \cite{li_narrow-band_2019} that, for the speech denoising task, the STFT coefficients of clean speech can be directly predicted by the multi-channel narrow-band network.
We follow and extend this principle in this work to the speech separation task, namely predicting the STFT coefficients of multiple speech signals. 
More specifically, the output $\widetilde{\mathbf{Y}}_{f}\in R^{2N\times T}$ consists of the real part and imaginary part of the (normalized) spatial image at the reference channel of $N$ speakers. 
\looseness=-1

The STFT coefficients at the original level of each separated speech signal, can be obtained from the network output $\widetilde{\mathbf{Y}}_f$ by first constructing the complex numbers, and then multiplying with the normalization factor: $\boldsymbol{\widehat{{\rm Y}}}_{f}^n = (\boldsymbol{\widetilde{{\rm Y}}}_{f}^{2n-1}+i\boldsymbol{\widetilde{{\rm Y}}}_{f}^{2n}) \overline{{\rm X}}_{f} \in \mathbb{C}^{1\times T}, n=1,\dots,N$.


As shown in Fig.\ref{fig1}, this paper uses a network which has two layers of bidirectional LSTM (BiLSTM) and one fully connected (FC) layer.
The network is shared by all frequencies, thus each frequency is processed independently.

\vspace*{-3mm}
\subsection{Full-band Permutation Invariant Training}

\vspace*{-1mm}
The training process of the proposed network has the label permutation problem, which can be solved by the PIT technique.
Applying PIT for each frequency separately, although the speech signals can be well separated at each frequency, it still suffers from the frequency permutation problem, as is for other narrow-band methods \cite{winter_map-based_2006, gannot_consolidated_2017, boeddecker_front-end_2018}.

To solve the frequency permutation problem, we propose a full-band PIT (fPIT) technique, which forces the network to produce predictions of all frequencies with an identical speaker label permutation. 
Specifically, the predictions at the same output position of all frequencies, 
i.e.\ $\widehat{\mathbf{Y}}^n = [\widehat{\mathbf{Y}}_0^n;\dots;\widehat{\mathbf{Y}}_{F-1}^n]\in \mathbb{C}^{F\times T}$,
are forced to belong to the same speaker, and binded together to form the complete spectra of this speaker.
Thus, the best permutation for the $N$ bindings can be regarded as the permutation for all frequencies.
The loss of all frequencies can then be calculated in a fPIT way:
\looseness=-1
\begin{small}
\begin{equation}
  \text{fPIT}(\boldsymbol{\rm \widehat{Y}}^{1},\ldots, \boldsymbol{\rm \widehat{Y}}^{N}, \boldsymbol{\rm {Y}}^{1},\ldots,\boldsymbol{\rm {Y}}^{N}) \\
  =
  \mathop{min}_{p\in \mathcal{P}}
  \frac{1}{N}
  \sum_n \mathcal{L}(
    \boldsymbol{{\rm {Y}}}^n
    ,
    \boldsymbol{{\rm \widehat{Y}}}^{p(n)}
  ) 
  \label{eq7}
\end{equation} 
\end{small}

\noindent where $\boldsymbol{{\rm {Y}}}^n \in \mathbb{C}^{F\times T}$ is a matrix consisting of all the STFT coefficients (of the spatial image at the reference channel) of the $n$-th speaker, i.e. ${\rm {Y}}_{f,t}^{n,r}, f=0,\dots,F-1; t=1,\dots,T$. 
$\mathcal{P}$ denotes the set of all possible permutations, and
$p$ denotes a permutation in $\mathcal{P}$ which maps labels of ground truth to labels of predictions.
$\mathcal{L}$ denotes a loss function.

The frequency binding in fPIT requires the network to not only separate the speech signals for each individual frequency,
but also output the predictions of all frequencies for one speaker at the same position,
even though the network processes frequencies separately.
The former task relies on learning narrow-band spatial information to discriminate multiple speakers.
The latter task possibly uses the (partial) narrow-band spatial information that are consistent along frequencies to determine the output position, such as the inter-channel cues related to the speaker direction.

fPIT uses the negative SI-SDR \cite{roux_sdr_2019} as the loss function:
\vspace*{-2mm}
\begin{equation}
   \mathcal{L}(\mathbf{Y}^n,\mathbf{\widehat{\mathbf{Y}}}^{p(n)}) 
   = -10 \log_{10}\frac{\left \| \alpha \boldsymbol{{\rm y}}^n \right \|^2}{\left \| \alpha \boldsymbol{{\rm y}}^n - \boldsymbol{{\rm \widehat{y}}}^{p(n)} \right \|^2}
  \label{eq8}
\end{equation}
where $\alpha=({\boldsymbol{{\rm \widehat{y}}}^{p(n)}})^T\boldsymbol{{\rm y}}^n/\left \| \boldsymbol{\rm y}^n \right \|^2$, $\boldsymbol{{\rm y}}^n$
and $\boldsymbol{{\rm \widehat{y}}}^{p(n)}$ are the inverse STFT of $\boldsymbol{{\rm Y}}^n$ and $\boldsymbol{{\rm \widehat{Y}}}^{p(n)}$, respectively.

\section{Experiments}
\label{sec:expset}

\subsection{Experimental Setup}

\textbf{Dataset.}
The proposed method is evaluated on a spatialized version of the WSJ0-2mix dataset \cite{hershey_deep_2016}.
The WSJ0-2mix dataset contains 20000, 5000 and 3000 speech pairs with varying lengths for training, validation and test respectively.
In this experiment, the speech pairs are overlapped in the manner used in \cite{luo_end--end_2020}, in which the tail part of one signal is overlapped with the head part of the other signal.
The overlap ratio is uniformly sampled in the range of [10\%, 100\%].
The resulting mixed utterances are all set to four-second long.
Room impulse responses are simulated using the gpuRIR package \cite{diaz-guerra_gpurir_2021}, which is a GPU based implementation of the image method \cite{allen_image_1979}.
The length, width and height of the simulated rooms are uniformly sampled in the range of [3 m, 8 m], [3 m, 8 m] and [3 m, 4 m], respectively.
The RT60 of each room is uniformly sampled in [0.1 s, 1.0 s].
An 8-channel circular microphone array with a radius of 5 cm is used.
The center of the array is randomly sampled in a square area (length is 1 m) in the center of the room at a height of 1.5 m.
The speaker locations are randomly sampled in the room with a height of 1.5 m and with the difference between the direction of two speakers randomly sampled from 0$^\circ$ to 180$^\circ$.
The speakers are located at least 0.5 m away from the walls.

\textbf{Training Configurations.}
\label{sec:modelcfg}
The sampling rate is 16 kHz.
STFT is performed using a hanning window of length 512 samples (32ms) with a hop size of 256 samples.
The numbers of hidden units of the first and second BiLSTM layers are set to 256 and 128 respectively.
The sizes of the input and output of the FC layer are 256 and 4 respectively.
The network is trained with 30 utterances per mini-batch, thus the batch size of frequencies is 7710 ($30 \times 257$).
The Adam \cite{kingma2015adam} optimizer is used with an initial learning rate of 0.001.
When the validation loss does not decrease in 10 consecutive epochs, the learning rate is halved until it reaches a given minimum of 0.0001.
Gradient clipping is applied with a threshold of 5.

\textbf{Performance Metrics and Comparison Methods.}
To evaluate the speech separation performance, three metrics are used: i) PESQ \cite{rix_perceptual_2001}, ii) SDR \cite{vincent_performance_2006}, iii) SI-SDR \cite{roux_sdr_2019}.
We compare the proposed method with two baselines: 
i) Oracle MVDR beamformer\footnote{https://github.com/Enny1991/beamformers}, for which the steering vector of desired speech and the covariance matrix of undesired signals are computed using the true desired speech and undesired signals, respectively; 
ii) The recently proposed FaSNet with the TAC mechanism \cite{luo_end--end_2020}, referred to as FaSNet-TAC.

\begin{table}[tbp]
    \caption{Speech separation results}
    \setlength{\tabcolsep}{1.0mm}{
    \vspace*{-3mm}
    \begin{tabular}{l|rrrr}
      Model & SDR & SI-SDR & NB-PESQ & WB-PESQ\\
      \hline
      Mixture & 0.18 & 0.00 & 2.05 & 1.6 \\
      Oracle MVDR & 12.19 & 11.70 & 3.21 & 2.68 \\
      FaSNet-TAC \cite{luo_end--end_2020} & 12.81 & 12.26 & 2.92 & 2.49 \\
      prop. with corr & 12.59 & 11.09 & 3.14 & 2.68 \\
      prop. & \textbf{13.89} & \textbf{13.26} & \textbf{3.31} & \textbf{2.87} \\
    \end{tabular}
    }
  \label{t1}
\end{table}

\subsection{Experimental Results}
\label{ssec:rd}

Table \ref{t1} shows the results. 
Compared with the oracle MVDR, FaSNet-TAC achieves higher SDR and SI-SDR scores, but lower NB-PESQ and WB-PESQ scores.
This means that FasNet-TAC is able to better separate the speech signals relying on the capability of the DNN, but the separated signals are possibly distorted and have a worse perceptual quality.
The proposed method achieves the best performance in terms of all metrics.
This indicates that narrow-band indeed involves rich information to discriminate between speakers, and the proposed network is able to exploit such information to well separate the signals.
The possible reason for the high PESQ scores and low speech distortion is that the proposed narrow-band method can better leverage the spatial constraints of speakers to preserve the signals, as is also done by MVDR.
The difference of perceptual quality between different methods is audible when listening to the separated sounds.
Some sound examples are available in our webpage \footnote{https://quancs.github.io/blog/nbss/}.

To evaluate the effectiveness of fPIT, the experiment using the correlation method
in \cite{sawada_robust_2004} for solving the frequency permutation problem is also conducted, in which fPIT is not used.
The speech separation results are shown in the line of “prop. with corr" in Table \ref{t1}.
Without fPIT, all performance measures significantly drop, which shows the effectiveness of fPIT for solving the frequency permutation problem.

\begin{figure}[t!]
    \centering
    \includegraphics[width=8cm]{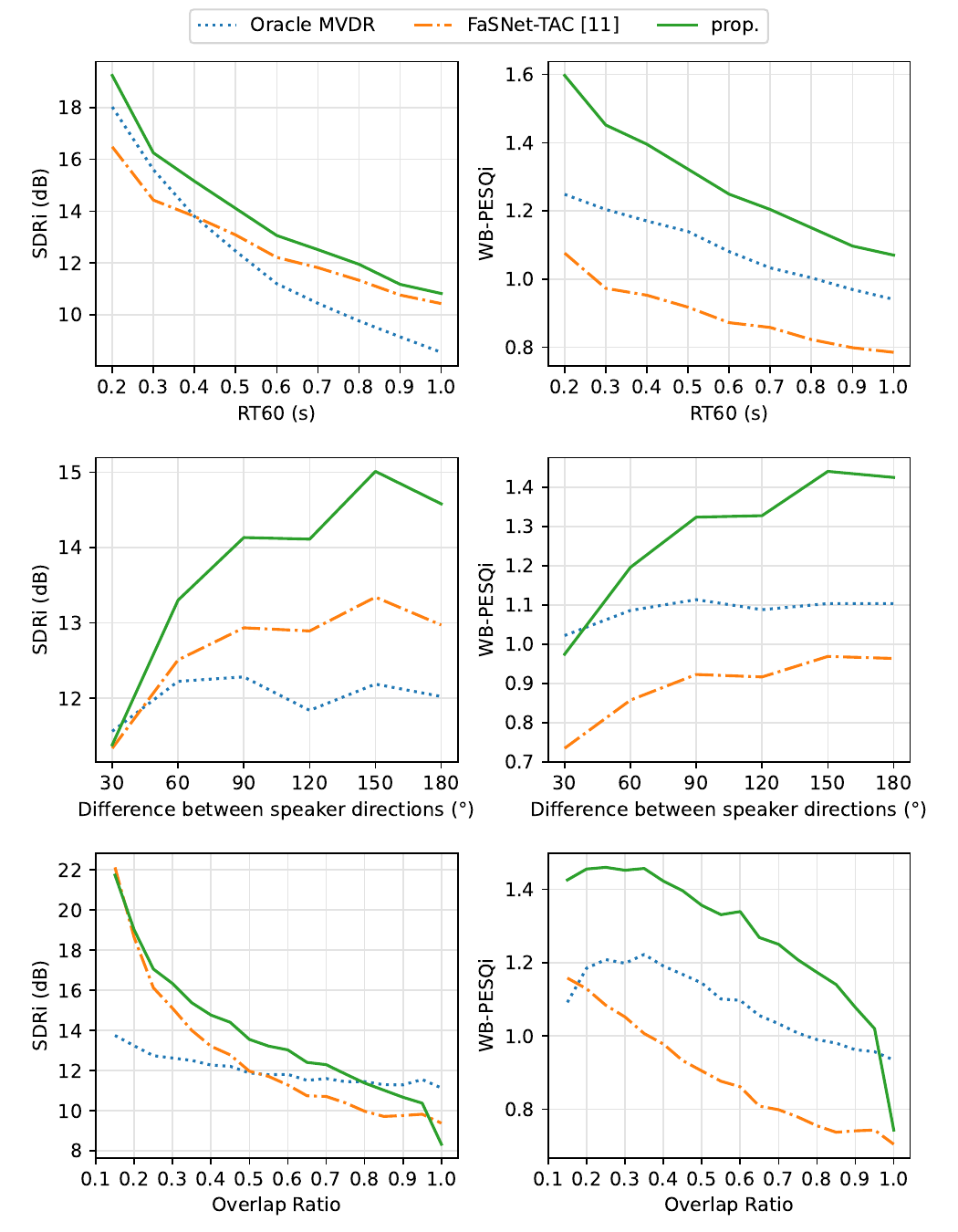}
    \vspace*{-3mm}
    \caption{The influence of (a) RT60, (b) difference between speaker directions, and (c) speech overlap ratio in terms of SDRi (SDR improvement) and WB-PESQi (WB-PESQ improvement)}
    \label{fig2}
\end{figure}

Fig. \ref{fig2} shows the performance improvement as a function of RT60, difference between speaker directions and speech overlap ratio. 
Not surprisingly, the performance of all three methods degrades with the increasing of RT60, since the reverberation distorts the spatial cues used for separation.
Regrading the difference between speaker directions, FaSNet-TAC and the proposed method achieve better performance when the difference increases.
A larger direction difference leads to a larger difference between the spatial cues of speakers, relying on which FaSNet-TAC and the proposed method separate the speakers.
By contrast, the influence of the direction difference for MVDR depends on array's beampattern. 
\looseness=-1

Along with the increase of speech overlap ratio, the performance of all the methods degrade, since non-overlapped signal is easier to separate than overlapped signal. 
Moreover, in the non-overlapped signals, each speaker presents solely, which provides some useful information, such as the speaker embeddings and the clean spatial cues, that can be used for separating the overlapped signals.
The performance of the proposed method significantly drops when the overlap ratio gets extremely high (higher than 95\%).
This indicates that the proposed method highly relies on the non-overlapped part to extract some information about the clean spatial cues, and much information can be provided even by a small portion (larger than 5\%) of non-overlapped signals.

\section{Conclusion}
\label{sec:concl}

In this paper, we proposed a narrow-band multi-channel speech separation network, which takes a single frequency band of STFT coefficients as input, and outputs the separated STFT coefficients of the same frequency.
The proposed network is trained with fPIT in an end-to-end way.
This work is the first deep learning based work exploring speech separation in a frequency by frequency fashion as many traditional speech separation methods do.
The experimental results show the superiority of the proposed method under most of the considered experimental conditions.
It is worth to note that the narrow-band information is hopefully complementary to the full-band information like spectral patterns of speech.
In the future, we will investigate how to fuse them for better speech separation.

\bibliographystyle{IEEEbib}
{\small
\bibliography{refs}}

\end{document}